\documentclass[prl,superscriptaddress,twocolumn,floatfix,showpacs,amsmath,amssymb]{revtex4}

\usepackage{graphicx}
\usepackage{dcolumn}
\usepackage{bm}
\usepackage[latin1]{inputenc}

\begin{document}

\title{Field and Temperature Dependence of the Superfluid Density in LaO$_{1-x}$F$_x$FeAs Superconductors: A Muon Spin Relaxation Study}

\author{H.~Luetkens}
\email{hubertus.luetkens@psi.ch}
\affiliation{Laboratory for Muon-Spin Spectroscopy, Paul Scherrer Institut,
CH-5232 Villigen PSI, Switzerland}
\author{H.-H.~Klauss}
\affiliation{IFP, TU Dresden, D-01069 Dresden, Germany}
\author{R.~Khasanov}
\affiliation{Laboratory for Muon-Spin Spectroscopy, Paul Scherrer Institut,
CH-5232 Villigen PSI, Switzerland}
\author{A.~Amato}
\affiliation{Laboratory for Muon-Spin Spectroscopy, Paul Scherrer Institut,
CH-5232 Villigen PSI, Switzerland}
\author{R.~Klingeler}
\affiliation{Institute for Solid State Research, IFW Dresden,
D-01171 Dresden, Germany}
\author{I.~Hellmann}
\affiliation{Institute for Solid State Research, IFW Dresden,
D-01171 Dresden, Germany}
\author{N.~Leps}
\affiliation{Institute for Solid State Research, IFW Dresden,
D-01171 Dresden, Germany}
\author{A.~Kondrat}
\affiliation{Institute for Solid State Research, IFW Dresden,
D-01171 Dresden, Germany}
\author{C.~Hess}
\affiliation{Institute for Solid State Research, IFW Dresden,
D-01171 Dresden, Germany}
\author{A.~K\"ohler}
\affiliation{Institute for Solid State Research, IFW Dresden,
D-01171 Dresden, Germany}
\author{G.~Behr}
\affiliation{Institute for Solid State Research, IFW Dresden,
D-01171 Dresden, Germany}
\author{J.~Werner}
\affiliation{Institute for Solid State Research, IFW Dresden,
D-01171 Dresden, Germany}
\author{B.~B\"uchner}
\affiliation{Institute for Solid State Research, IFW Dresden, D-01171 Dresden, Germany}


\date{\today}

\begin{abstract}
We present zero field and transverse field $\mu$SR
experiments on the recently discovered electron doped Fe-based
superconductor LaO$_{1-x}$F$_x$FeAs. The zero field experiments on
underdoped ($x$=0.075) and optimally doped ($x$=0.1) samples rule out any static magnetic order above 1.6~K in these superconducting samples. From transverse field experiments in the vortex phase
we deduce the temperature and field dependence of the superfluid
density. Whereas the temperature dependence is consistent with a
weak coupling BCS s-wave or a dirty d-wave gap function scenario, the field dependence strongly evidences unconventional superconductivity. We
obtain the in-plane penetration depth of  $\lambda_\mathrm{ab} (0) =
254(2)$~nm for LaO$_{0.9}$F$_{0.1}$FeAs and $\lambda_\mathrm{ab} (0)
= 364(8)$~nm for LaO$_{0.925}$F$_{0.075}$FeAs. Further evidence for unconventional superconductivity is provided by the ratio of $T_\mathrm{c}$ versus the
superfluid density, which is close to the Uemura line of hole doped
high-$T_\mathrm{c}$ cuprates. 
\end{abstract}



\maketitle


The ongoing search for new superconductors has recently yielded a new family of Fe-based compounds composed of alternating $\rm La_2O_{2-x}F_x$ and $\rm Fe_2As_2$ layers with
transition temperatures $T_\mathrm{c}$ up to 28~K \cite{Kamihara08}. By replacing
La with other rare earths, $T_\mathrm{c}$ can be raised to above 50~K \cite{Chen08_XH-arXiv,Chen08_GFb-arXiv,Ren08a-arXiv,Ren08b-arXiv,Cheng08-arXiv,Ren08c-arXiv}, and thus the first non-copper-oxide superconductor with $T_\mathrm{c}$ exceeding 50~K has emerged. Both, recent experimental findings and theoretical treatments \cite{Cao08-arXiv,Eschrig-arXiv} indicate unconventional multiband superconductivity in the layers of paramagnetic Fe ions, which would normally destroy superconductivity in traditional s-wave superconductors.
Point contact tunneling spectroscopy
\cite{Shan08-arXiv}, specific heat \cite{Mu08-arXiv} and
magnetization measurements \cite{Ren_C08-arXiv} point to nodal order
parameters. High magnetic field experiments yielded evidence for
two-band effects \cite{Hunte08-arXiv}.
Various scenarios for superconductivity have also been discussed theoretically and different pairing symmetries of the
superconducting ground state including spin-triplet p-wave pairing
have been proposed
\cite{Xu08-arXiv,Mazin08-arXiv,Kuroki08-arXiv,Dai08-arXiv,Han08-arXiv,Boeri08-arXiv,Lee08-arXiv}.
Intriguingly, there is evidence for a close interplay between superconductivity and magnetism as it is well established for other unconventional superconductors. A commensurate spin-density wave (SDW) has been observed below 150~K in the undoped compound \cite{Ma08-arXiv,Dong08-arXiv,Cruz08-arXiv,McGuire08-arXiv}, and a recent theoretical work suggests that fluctuations associated with a magnetic quantum critical point are essential for superconductivity in the F-doped system \cite{Giovannetti08-arXiv}.

In this Letter, we report zero field (ZF) and high transverse field (TF) muon
spin relaxation measurements ($\mu$SR) on  polycrystalline samples
of LaO$_{1-x}$F$_x$FeAs with $x$=0.075 and 0.10. Our ZF-$\mu$SR
experiments show that no static magnetic correlations are present down
to 1.6~K. Hence, the spin-density state is completely suppressed upon F-doping.
Properties of the superconducting state are determined by the TF-$\mu$SR measurements:
The weak temperature dependence of the superfluid density
below $T_\mathrm{c}/3$ is consistent with both an s-wave and a dirty d-wave scenario. However, the field dependence of the static line-width contradicts a standard BCS s-type gap function. In addition, our data provide the first quantitative measurements of the in-plane penetration depths, which amount to several hundred nanometers. This implies a dilute superfluid in LaO$_{1-x}$F$_x$FeAs and, remarkably, the data for the Fe-based superconductors are very close to the Uemura line of the hole doped high-$T_\mathrm{c}$ cuprates.


Polycrystalline samples of LaO$_{1-x}$F$_x$FeAs ($x$= 0.075, 0.1) were
prepared
by using a two-step
solid state reaction method, similar to that described by Zhu et al.
\cite{Zhu_X08-arXiv}, and annealed in vacuum. The samples consists of 1 to 100~$\mu$m sized grains of
LaO$_{1-x}$F$_x$FeAs. The crystal structure and the composition were investigated by powder X-ray diffraction and wavelength-dispersive X-ray spectroscopy (WDX).

\begin{figure}[htbp]
\center{\includegraphics[width=0.9\columnwidth,angle=0,clip]{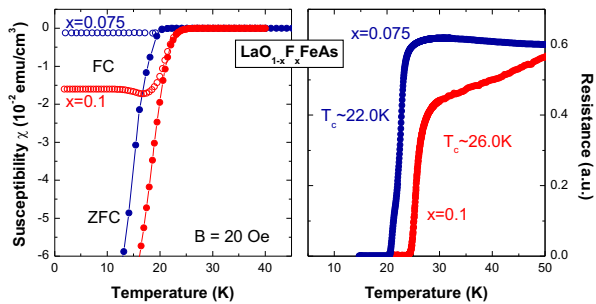}}
    \caption[]{Right: temperature dependence of the resistance of LaO$_{1-x}$F$_{x}$FeAs at $x=0.1$ and $x=0.075$ in the vicinity of $T_\mathrm{c}$. Left: field cooled and zero field cooled
magnetic susceptibility for $x=0.1$ and $x=0.075$.}
\label{chi_rho}
\end{figure}

In order to characterize the superconducting properties, zero field cooled (shielding signal) and field cooled (Meissner
signal) magnetic susceptibility in external fields $H=10~$Oe$\dots50$~kOe have been
measured using a SQUID magnetometer.
The resistance has been measured with a standard 4-point geometry employing an alternating DC current.
Critical temperatures of $T_\mathrm{c}\approx 26.0$~K and $T_\mathrm{c}\approx 22$~K for $x=0.1$ and $x=0.075$, respectively, are extracted from these measurements (cf. Fig.~\ref{chi_rho}).



\begin{figure}[htbp]
\center{\includegraphics[width=0.9\columnwidth,angle=0,clip]{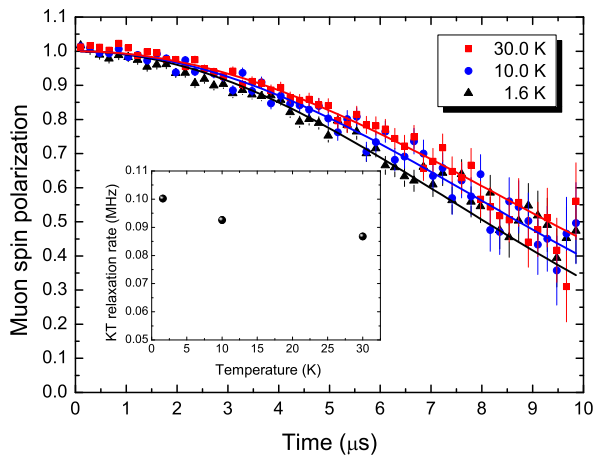}} \caption[]{Zero field $\mu$SR spectra of LaO$_{0.9}$F$_{0.1}$FeAs for 1.6, 10, and 30~K. The inset shows the Gaussian Kubo-Toyabe relaxation rate as a function of temperature.}
\label{ZF-spectra}
\end{figure}

In Fig.~\ref{ZF-spectra} we show representative ZF-$\mu$SR data for $x=0.1$ at 1.6, 10, and 30~K. At all temperatures a weak Gaussian Kubo-Toyabe-like \cite{Hayano79} (KT) decay  of the muon spin polarization is observed. 
The relaxation rates are very small and can be traced back to the tiny magnetic fields originating from nuclear moments. This implies that we can rule out any static SDW magnetism with considerable magnetic moments, for both the optimally and the underdoped samples. Thus, our data are in striking contradiction with the prediction in Ref.~\onlinecite{Cao08-arXiv}. Instead, our
measurements show that doping the compound LaOFeAs with electrons suppresses the SDW instability, and simultaneously promotes the
superconductivity as the new ground state. This indicates that the superconducting phase is close to a quantum critical point related to the magnetic instability.

Such a proximity of magnetism and superconductivity is also suggested by several aspects of our $\mu$SR-data. The first weak hint is a slight increase of the relaxation rate below $T_\mathrm{c}$ (cf. inset Fig.~\ref{ZF-spectra}). Further indications arise from the high-TF measurements on LaO$_{0.925}$F$_{0.075}$FeAs, as will be discussed below.

\begin{figure}[htbp]
\center{\includegraphics[width=0.9\columnwidth,angle=0,clip]{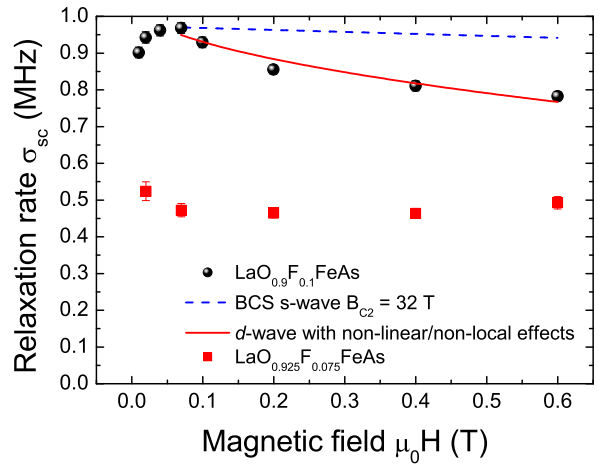}}
    \caption[]{Field dependence $\sigma_\mathrm{sc}$ at 1.6~K for LaO$_{0.925}$F$_{0.075}$FeAs and LaO$_{0.9}$F$_{0.1}$FeAs. The dashed line is the expected behavior for an s-wave BCS superconductor with $\mu_0H_\mathrm{c2}=32$~T according to Eq.~\ref{eq:sigma_vs_h}. The solid line is a fit of the data with Eq.~\ref{nonlinear} indicative for nodes in the gap function.}
\label{B-scan}
\end{figure}

\begin{figure}[htbp]
\center{\includegraphics[width=0.9\columnwidth,angle=0,clip]{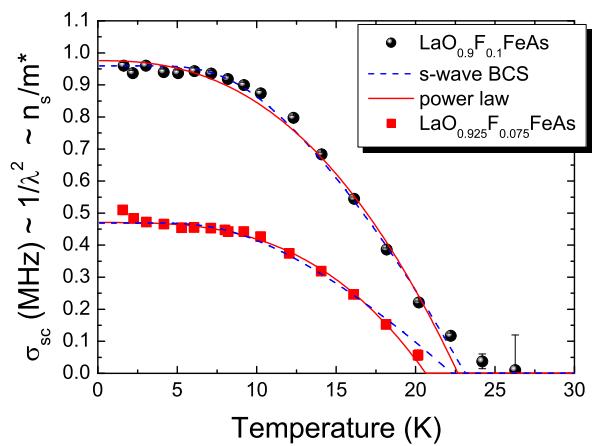}}
    \caption[]{Temperature dependence of $\sigma_\mathrm{sc}$ measured in a field of $\mu_0H=0.07$~T for LaO$_{0.925}$F$_{0.075}$FeAs and LaO$_{0.9}$F$_{0.1}$FeAs. The $\sigma_\mathrm{sc}(T)$ were fitted using a standard BCS curve and a power law $1-(T/T_\mathrm{c})^{n}$. In case of LaO$_{0.925}$F$_{0.075}$FeAs the points below 4~K were omitted in the fit.}
\label{fig:sigma_vs_T}
\end{figure}

The results of the analysis of the magnetic field dependent and temperature dependent TF-$\mu$SR measured in the vortex phase $H>H_\mathrm{c1} \approx 40$~Oe are displayed in Figures~\ref{B-scan} and \ref{fig:sigma_vs_T}.
From the muon spin polarization $P(t)$ we determined the Gaussian relaxation rate $\sigma$ which is the sum of a nuclear
contribution $\sigma_{nm}$ and a contribution $\sigma_{sc}$
proportional to the second moment of the magnetic field distribution
of the vortex lattice, i.e. $\sigma^{2}=\sigma_{sc}^2 + \sigma_{nm}^2$.
In order to extract the $\sigma_{sc}$ which measures the superfluid density, i.e.  $\sigma_{sc}^2 \propto 1/\lambda^2 \propto n_\mathrm{s}/m^*$ \cite{Brandt88}, we determined the small nuclear relaxation rate $\sigma_\mathrm{nm} = 0.07(1)$~MHz at 30~K.

Figure~\ref{B-scan} shows the obtained field dependence of $\sigma_\mathrm{sc}$ for both
LaO$_{1-x}$F$_x$FeAs samples at 1.6~K. Each measurement was performed after
cooling the sample in the field from above $T_\mathrm{c}$. 

We restrict our discussion of $\sigma_\mathrm{sc}(H)$ to the optimally doped sample since the data for $x=0.075$ are clearly influenced by additional contributions. In particular, after a small decrease we find that $\sigma_\mathrm{sc}(H)$ starts to increase with increasing field, which is incompatible with the suppression of the superfluid density in high magnetic fields. Analogous behavior has been observed in 
high-$T_\mathrm{c}$ cuprates \cite{Savici05,Khasanov07}, where an external field can promote the magnetic correlations leading to spurious magnetism in the otherwise superconducting sample. The upturn of $\sigma_\mathrm{sc}$ as a function of magnetic field is therefore most likely produced by an additional magnetic contribution and does not reflect a field dependence of the superfluid density.

In contrast, the data for $x=0.1$ are compatible with a $\sigma_\mathrm{sc}$, that is purely originating from the second moment of the field distribution of vortex lattice. At low fields a maximum in $\sigma_\mathrm{sc}(H)$ is observed followed by a
decrease of the relaxation rate up to the highest fields. At first glance this appears to be consistent with a BCS s-wave superconductor. However, a quantitative analysis taking into account the large critical fields reveals strong discrepancies.

In a conventional s-wave superconductor the penetration depth $\lambda$ is field
independent and $\sigma_\mathrm{sc}$ increases with increasing
magnetic field up to $H\simeq 2H_\mathrm{c1}$. At higher fields
a weak field dependence according to Eq.~\ref{eq:sigma_vs_h}
is expected in an ideal triangular vortex lattice \cite{Brandt88}:
\begin{equation}
 \label{eq:sigma_vs_h}
\sigma_{sc}[\mu {\rm s}^{-1}]=4.83\times 10^4 (1-h)
  [1+3.9(1-h)^2]^{1/2}
 \lambda^{-2}[{\rm nm}] \; .
\end{equation}
Here, $h=H/H_\mathrm{c2}$ and $H_\mathrm{c2}$ is the upper critical
field which has been reported to be as large as $32\dots65$~T in optimally doped LaO$_{1-x}$F$_x$FeAs \cite{Sefat08-arXiv,Zhu_X08-arXiv,Hunte08-arXiv}. We note that high field studies on our sample corroborate these large values \cite{Fuchs_unpub}.
Taking these critical fields, the theoretical BCS behavior has been calculated. It is shown by the dashed line in Fig.~\ref{B-scan}. The
decrease of $\sigma_\mathrm{sc}$ at $\mu_0H\gtrsim 0.1$~T is in striking contradiction with conventional s-wave BCS behavior as given by
Eq.~\ref{eq:sigma_vs_h}. It is only reproduced if an unrealistically small  $\mu_0H_\mathrm{c2}=5$~T is assumed. 

For unconventional superconductors with nodes in the gap, $\lambda$
depends on the field. This leads to a decrease of
$\sigma_\mathrm{sc}(H)$ for $H>2H_\mathrm{c1}$, which is generally observed for various
high-$T_\mathrm{c}$ superconductors \cite{Sonier00}. This
observation is nicely described by theories taking non-local/non-linear effects into account. In this case, the field dependence
of the superfluid density can be described by
\cite{Won01,Vekhter99}:
\begin{equation}
\frac{\lambda^{-2}(H)}{\lambda^{-2}(H=0)}=
\frac{\sigma_\mathrm{sc}(H)}{\sigma_\mathrm{sc}(H=0)} = 1 - K \sqrt{H}\, ,
\label{nonlinear}
\end{equation}
where $K$ is the parameter depending on the
strength of the non-linear effects.
Applying this model, our data are well described by Eq.~\ref{nonlinear} with $K=0.0034$~Oe$^{-0.5}$ indicative for superconductivity with nodes in the gap function \footnote{We note that a maximum in $\sigma_\mathrm{sc}(H)$ may be attributed to vortex lattice distortions caused by strong pinning. However, in
our samples pinning is weak as signaled by the field
cooled susceptibility shown in Fig.~\ref{chi_rho}.}.



It is well established that nodes in the gap function also influence the temperature dependence of the superfluid density $n_\mathrm{s}/m^* \propto 1/\lambda^2$.
This temperature dependence $n_\mathrm{s}(T)$ is directly obtained from the temperature dependent TF-$\mu$SR measurements. To ensure an accurate determination of
$\lambda(T)$ it is mandatory to measure slightly above the maximum of
$\sigma_\mathrm{sc}(H)$ where Eq.~\ref{eq:sigma_vs_h} is valid to
determine the absolute value of $\lambda$. Therefore, the
measurements were done with an external field of 700~G. The results for $\sigma_\mathrm{sc}(T)$ are shown in
Figure~\ref{fig:sigma_vs_T}.

Surprisingly, and in contrast to our conclusions from the field dependence, the results follow the s-wave  weak
coupling BCS temperature dependence as shown by the dashed
curve. 
However, the data are also
reasonably well described by a power law $1-(T/T_\mathrm{c})^{2.45(4)}$, which is similar to the prediction for the
dirty limit d-wave model $1-(T/T_\mathrm{c})^2$ \cite{Hirschfeld94}.
Other gap symmetries such as a clean limit d-wave or a non-monotonic d-wave have been tested also, but were found to be inconsistent with the data. In particular, it is difficult to account for the very weak temperature dependence of $\sigma_\mathrm{sc}$ at low $T$ within such models.
Considering both field and temperature dependence of $\sigma_\mathrm{sc}$, the dirty-limit d-wave model appears to be most compatible with the data. It it important to note, that we observe a strong reduction of the high-TF Knight shift below T$_\mathrm{c}$. This points to
significant reduction of the spin susceptibility in the
superconducting state which excludes triplet pairing.

The temperature dependence for the underdoped compound LaO$_{0.925}$F$_{0.075}$FeAs resembles that of the optimum doped sample, with the difference that an upturn of $\sigma_{sc}$ is observed below 4~K. When measuring $\sigma_{sc}(T)$ in a higher field of 0.6~T, the upturn of $\sigma_{sc}(T)$ is already observed at higher temperatures $T\lesssim 7$~K. In other words, the upturn is not due to the temperature dependence of the superfluid density but a further hint to the proximity to magnetism which we have discussed in the context of the field dependent measurements shown in Fig.~\ref{B-scan}. 

We now turn our discussion on $\lambda (T=0)$ and the in-plane
penetration depth $\lambda_{ab}$. For LaO$_{0.9}$F$_{0.1}$FeAs we
determine a zero temperature relaxation rate $\sigma_\mathrm{sc}(0)
= 0.965(10)$~MHz. Using Eq.~\ref{eq:sigma_vs_h} the effective
penetration depth can be calculated to $\lambda (0) = 333(2)$~nm.
For LaO$_{0.925}$F$_{0.075}$FeAs we analogously determined the low
temperature relaxation rate to $\sigma_\mathrm{sc}(0) =
0.47(2)$~MHz, yielding a penetration depth of $\lambda (0) =
477(10)$~nm.
For powdered samples the experimentally extracted $\lambda$ is a
geometrically averaged penetration depth. However, in
LaO$_{1-x}$F$_x$FeAs a rather large anisotropy can be expected from recent band structure calculations predicting e.g. a resistivity
ratio of $\rho_\mathrm{c}/\rho_\mathrm{ab} \approx 10-15$ \cite{Singh08-arXiv}.
In our TF-$\mu$SR measurements a large anisotropy of $\lambda$ is
confirmed by the observation of a symmetric Gaussian shape of the
field distribution $p(B)$. In contrast, for an isotropic
superconductor a typical nonsymmetric shape of $p(B)$ with a van Hove
singularity is found \cite{Sonier00}. It has been shown
\cite{Fesenko91} that for large anisotropies the measured effective
penetration depth $\lambda$ becomes independent on the actual
anisotropy and is solely determined by the in-plane penetration
depth $\lambda_\mathrm{ab}$ and can be expressed by $\lambda = 1.31
\lambda_\mathrm{ab}$. Therefore we obtain the in-plane penetration
depth of  $\lambda_\mathrm{ab} (0) = 254(2)$~nm for
LaO$_{0.9}$F$_{0.1}$FeAs and $\lambda_\mathrm{ab} (0) = 364(8)$~nm
for LaO$_{0.925}$F$_{0.075}$FeAs, respectively.

\begin{figure}[htbp]
\center{\includegraphics[width=0.9\columnwidth,angle=0,clip]{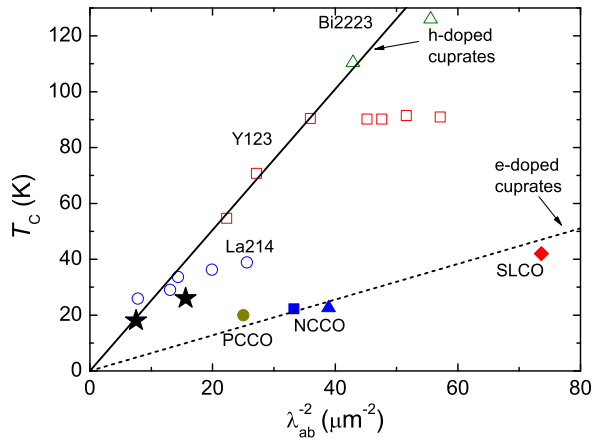}}
    \caption[]{Uemura plot for hole and electron doped high $T_\mathrm{c}$ cuprates.
    Points for the cuprates are taken from \cite{Uemura91,Nugroho99,Homes97,Shengelaya05,Homes06}.
    The stars are showing the data for LaO$_{1-x}$F$_x$FeAs obtained in this work.}
\label{Uemura-plot}
\end{figure}

In Fig.~\ref{Uemura-plot} we display our data in the so-called Uemura plot, which nicely demonstrates the linear relation of superfluid density and $T_\mathrm{c}$ for under and optimally doped superconductors \cite{Uemura89}. We compare our data to the cuprate family. The data for LaO$_{1-x}$F$_x$FeAs are close to the Uemura line for hole doped cuprates, indicating that the superfluid is also very dilute in the oxypnictides. This observation is in accordance with its small normal state charge carrier density \cite{Chen08_GFa-arXiv,Yang08-arXiv,Zhu_X08-arXiv} and provides further evidence for the unconventional superconductivity in LaO$_{1-x}$F$_x$FeAs.

In conclusion, we have performed zero field and high transverse field muon
spin relaxation measurements on polycrystalline samples
of LaO$_{1-x}$F$_x$FeAs with $x$=0.075 and 0.10. Our ZF-$\mu$SR
experiments show that the spin-density state is completely suppressed upon F-doping, which suggests a close proximity of the superconducting phase to a magnetic quantum critical point. TF-$\mu$SR measurements reveal a weak temperature dependence of the superfluid density
below $T_\mathrm{c}/3$, which is consistent with both an s-wave and a dirty d-wave scenario. However, the field dependence supports an unconventional order parameter, but a clean-limit d-wave as well as triplet pairing can be clearly excluded by our data.
We provide quantitative measurements of the in-plane penetration depths, which amount   to $\lambda_\mathrm{ab} (0) =
254(2)$~nm and $\lambda_\mathrm{ab} (0)
= 364(8)$~nm for $x=0.1$ and $x=0.075$, respectively. Hence, the superfluid in LaO$_{1-x}$F$_x$FeAs is dilute and, remarkably, the data for the Fe-based superconductors are very close to the Uemura line of the hole doped high-$T_\mathrm{c}$ cuprates.


%

\begin{acknowledgments}
This work was performed at the Swiss Muon Source, Paul Scherrer Institut, Villigen,
Switzerland. We thank M. Deutschmann, S. M\"uller-Litvanyi, R. M\"uller and J. Hamann-Borrero for experimental support in preparation and characterization of the samples.
The work at the IFW Dresden has been supported by the Deutsche Forschungsgemeinschaft through Forschergruppe FOR 538.
\end{acknowledgments}


\end{document}